\documentstyle[12pt, slashbox]{article}
\def\BorL{b }	
\makeatletter
\newif\ifpr@pstyle \pr@pstylefalse
\newif\ifnons@qeq  \nons@qeqfalse
\def\bigpage{
	\setlength{\topmargin}{-.5in}
	\setlength{\oddsidemargin}{.5pc}
	\setlength{\evensidemargin}{.5pc}
	\setlength{\textwidth}{35pc}
	\setlength{\textheight}{56pc}
	\setlength{\parskip}{6pt plus 2pt minus 1pt}
	\newlength{\paperbaselineskip}
	\setlength{\paperbaselineskip}{20pt plus 0.2pt minus 0.1pt}
	\def\@oddfoot{\hfil -- \thepage~--\hfil}
	\let\@evenfoot\@oddfoot
        \def\thesection{\arabic{section}.}
        \def\thesubsection{\thesection\arabic{subsection}}
        \def\@ourappendix{\par\setcounter{section}{0}
                      \setcounter{subsection}{0}
                      \def\thesection{\Alph{section}.}
                      \ifnons@qeq
                      \def\theequation{\Alph{section}.\arabic{equation}}\fi}
        \def\appendix{\@ourappendix}
        \def\section{\@startsection {section}{1}%
            {\z@}{5ex plus .2ex minus .4ex}%
            {1.5ex plus.4ex minus .1ex}%
            {\centering\ifpr@pstyle\else\ifx\undefined\reset@font\else%
             \reset@font\fi\large\fi\bf}}
        \def\subsection{\@startsection{subsection}%
            {2}{\z@}{3.25ex plus .4ex minus .4ex}%
            {1ex plus .2ex}{\bf}}
}
\bigpage
\newfont{\fourteencp}{cmcsc10 scaled\magstep2}
\newfont{\titlefont}{cmbx10 scaled\magstep2}
\newfont{\authorfont}{cmcsc10 scaled\magstep1}
\newfont{\fourteenmib}{cmmib10 scaled\magstep2}
	\skewchar\fourteenmib='177
\newfont{\elevenmib}{cmmib10 scaled\magstephalf}
	\skewchar\elevenmib='177
\newfont{\ninemib}{cmmib9} \skewchar\ninemib='177
\makeatletter
\newcommand\nonsequentialeqnum{
        \nons@qeqtrue
	\@addtoreset{equation}{section}
	\def\theequation{\arabic{section}.\arabic{equation}}}
\newif\ifp@bblock  \p@bblocktrue
\newcommand\nopubblock{\p@bblockfalse}
\newcommand\topspace{\hrule height 0pt depth 0pt \vskip}
\newcommand\p@bblock{\begingroup \tabskip=\hsize minus \hsize
	\baselineskip=1.5\ht\strutbox \topspace-2\baselineskip
	\halign to\hsize{\strut ##\hfil\tabskip=0pt\crcr
	\the\Pubnum\crcr\the\date\crcr}\endgroup}
\newcommand\YUKAWAmark{\hbox{
        \ifpr@pstyle\ninemib\else\elevenmib\fi
        Yukawa\hskip1mm Institute\hskip1mm Kyoto \hfill}}
\newtoks\date
\newtoks\Pubnum
\let\pubnum=\Pubnum
\Pubnum={}
\date={\today}
\newcommand{\frontpageskip}{\vspace{12pt plus .5fil minus 2pt}}
\def\@authoraddress{} \def\@title{}
\def\title#1{\gdef\@title{\frontpageskip
	\begin{center}{\titlefont #1}\end{center}\par}}
\def\@author#1{\frontpageskip\par\begin{center}{\authorfont #1}
	\end{center}
	\nobreak}
\def\author#1{\expandafter\def\expandafter\@authoraddress\expandafter
    {\@authoraddress{\@author{#1}}}}
\def\andauthor#1{\expandafter\def\expandafter\@authoraddress\expandafter
    {\@authoraddress{\frontpageskip\centerline{and}\@author{#1}}}}
\def\authors#1{\expandafter\def\expandafter\@authoraddress\expandafter
    {\@authoraddress{\frontpageskip\noindent #1}}}
\def\@address#1{\par\begin{center}{\sl #1}\end{center}\par}
\def\address#1{\expandafter\def\expandafter\@authoraddress\expandafter
    {\@authoraddress{\@address{#1}}}}
\def\andaddress#1{\expandafter\def\expandafter%
    \@authoraddress\expandafter
    {\@authoraddress{\par\centerline{\sl and}\@address{#1}}}}
\renewcommand{\thanks}[1]{\footnote{#1}}
\def\maketitle{\par
  \begingroup
       \def\thefootnote{\fnsymbol{footnote}}
	\thispagestyle{empty}
        \baselineskip=\paperbaselineskip
	\@maketitle
	\endgroup
	\setcounter{footnote}{0}
	\let\maketitle\relax \let\@maketitle\relax
	\let\@thanks\relax \let\@title\relax
	\let\@title\relax \let\@authoraddress\relax
	\let\thanks\relax}
\def\@maketitle{%
        \ifpr@pstyle\vspace{-1.0cm}\else\vspace{-1.7cm}\fi
	\YUKAWAmark\vskip0.6cm
	\ifp@bblock\p@bblock \else\hrule height 0pt \relax \fi
	\@title
	\@authoraddress
	}
\renewcommand{\abstract}{\par\frontpageskip\centerline{
             \ifpr@pstyle\twelvecp\else\fourteencp\fi Abstract}
	\vspace{8pt plus 3pt minus 3pt}}
\@addtoreset{equation}{section}
\def\theequation{\arabic{section}.\arabic{equation}}
\makeatother
\def\bigmode{b }
\ifx\BorL\undefined\read-1 to\BorL\fi
\makeatletter
\def\doublepage{
        \twocolumn
        
        \pr@pstyletrue
        \sloppy
        \flushbottom
        \setlength{\topmargin}{-0.95in}
        \setlength{\headsep}{20pt}
        \setlength{\headheight}{10pt}
        \hoffset=-0.35in
        \leftmargini 2em
        \leftmarginv .5em
        \leftmarginvi .5em
        \marginparwidth 48pt
        \marginparsep 10pt
        \setlength{\columnsep}{0.7truein}
        \setlength{\textwidth}{10.5truein}
        \setlength{\textheight}{7truein}
        \setlength{\oddsidemargin}{0.0truein}
        \setlength{\evensidemargin}{0.0truein}
        \multiply\paperbaselineskip by 4
                   \divide\paperbaselineskip by 5
        \multiply\footskip by 4 \divide\footskip by 5
        \setlength{\parskip}{4pt plus 1.5pt minus 1pt}
        \newlength{\halfwidth}
        \halfwidth=\textwidth\advance\halfwidth by -\columnsep
                         \divide\halfwidth by 2
        \newfont{\twelvemib}{cmmib10 scaled\magstep1}
                 \skewchar\twelvemib='177
        \newfont{\tenmib}{cmmib10}
                 \skewchar\tenmib='177
        \newfont{\twelvecp}{cmcsc10 scaled\magstep1}
        \def\pagebox{\hbox to \halfwidth{\hfil  -- \thepage~--\hfil}}
        \def\@oddfoot{\pagebox\hfil\addtocounter{page}{1}\pagebox}
        \let\@evenfoot\@oddfoot
        \def\ps@empty{\let\@mkboth\@gobbletwo\let\@oddhead\@empty
               \def\@oddfoot{\hbox to \halfwidth{\hfil ~~~~~~~}\hfil
               \addtocounter{page}{1}\pagebox}
                \let\@evenhead\@empty\let\@evenfoot\@oddfoot}
        \def\appendix{\@ourappendix}
        \def\section{\@startsection {section}{1}%
            {\z@}{5ex plus .2ex minus .4ex}%
            {1.5ex plus.4ex minus .1ex}%
            {\centering\ifpr@pstyle\else\reset@font\large\fi\bf}}
        \def\subsection{\@startsection{subsection}%
            {2}{\z@}{3.25ex plus .4ex minus .4ex}%
            {1ex plus .2ex}{\bf}}
}
\ifx\BorL\bigmode
\else
        \input art10.sty
        \doublepage
\fi
\makeatother
\makeatletter
\newif\ifepsfloaded
\newif\iffigureexists

\ifeof 1 \epsfloadedfalse \else \epsfloadedtrue \fi
\ifepsfloaded \input epsf \fi

\def\checkex#1 {\relax
    \openin 1 #1
    \ifeof 1 \figureexistsfalse
    \else \figureexiststrue
    \fi \closein 1 }
\def\figinsertraw#1#2{
   \ifepsfloaded
       \checkex #1
       \iffigureexists
           \immediate\write16{(#1)}
           #2
       \else
           \immediate\write16{(#1 NOT FOUND!)}
           \vbox to 2in{\hbox to 2in {\hss} \vss}
       \fi
   \else
       \immediate\write16{(NOT inputting #1; no epsf.tex)}
       \vbox to 2in{\hbox to 2in {\hss} \vss}
   \fi}
\newcommand{\reduceland}[2]{\dimen@=#1
     \ifpr@pstyle\multiply\dimen@ by 4\divide\dimen@ by 5\fi
     \edef#2{\dimen@}}
\def\F@gin#1#2#3#4{
  \ifepsfloaded
    \checkex #1
    \iffigureexists
        \immediate\write16{(#1)}
        \begin{figure}
        \ifdim#2>\z@\reduceland{#2}{\dimen@ii}\epsfxsize=\dimen@ii\fi
        \ifdim#3>\z@\reduceland{#3}{\dimen@ii}\epsfysize=\dimen@ii\fi
        \centerline{\epsfbox{#1}}
        {#4} \end{figure}
    \else
        \immediate\write16{(#1 NOT FOUND!)}
        \begin{figure}
        \ifdim#2>\z@\reduceland{#2}{\dimen@ii}\else\dimen@ii=2in\fi
        \ifdim#3>\z@\reduceland{#3}{\dimen255}\else\dimen255=2in\fi
        \centerline{\framebox[\dimen@ii]{\rule{0pt}{\dimen255}#1}}
        {#4} \end{figure}
    \fi
  \else
    \immediate\write16{(NOT inputting #1; no epsf.tex)}
    \begin{figure}
    \centerline{\framebox[2in]{\rule{0pt}{2in}#1}}
    #4\end{figure}
  \fi}
\def\figinsertx#1#2#3{\F@gin{#1}{#2}{0pt}{#3}}
\def\figinserty#1#2#3{\F@gin{#1}{0pt}{#2}{#3}}
\def\figinsert#1#2{\F@gin{#1}{0pt}{0pt}{#2}}
\makeatother
\begin{document}  %
%
\pubnum{YITP-04-56\cr OIQP-04-3}
\date{Oct. 2004}

\title{Dirac Sea for Bosons I\\
--- Formulation of Negative Energy Sea for Bosons ---
\thanks{This paper is the first part of the revised version of ref. [2].}
}

\author{Holger B. Nielsen
}
\address{Niels Bohr Institute, \\
	University of Copenhagen, 17 Blegdamsvej, \\
	Copenhagen $\o$, DK 2100, Denmark\\
}

\andauthor{Masao Ninomiya
\thanks{Also, Okayama Institute for Quantum Physics, Kyo-yama 1-9-1,
Okayama City 700-0015, Japan
}
}
\address{Yukawa Institute for Theoretical Physics\\
        Kyoto University,~Sakyo-ku,~Kyoto 606-8502,~Japan\\}

\maketitle


\begin{abstract}
It is proposed to make formulation of second quantizing a bosonic
theory by generalizing the method of filling the Dirac's negative energy
sea for fermions.
We interpret that the correct vacuum for the bosonic theory is
obtained  by adding minus one boson to each single particle negative
energy states while the positive energy states are empty.
The boson states are divided into two sectors ;
the usual positive sector with positive and zero numbers of bosons and
the negative sector with negative number of bosons.
Once it comes into the negative sector it cannot return to the usual
positive sector by ordinary interaction due to a barrier.

It is suggested to use as a playground model in which the filling of
empty fermion Dirac sea and the removal of boson from the
negative energy states are not yet performed.
We put forward such a naive vacuum world
in the present paper.
The successive paper[1] will concern various properties: Analyticity
of the wave functions, interaction and a CPT-like Theorem in the naive
vacuum world.

\end{abstract}


\section{Introduction}

There has been a wellknown method, though not popular nowadays, to second
quantize relativistic fermion by imagining that there is a priori so
called naive vacuum in which there is no, neither positive energy nor
negative energy, fermion present.
However this vacuum is unstable and the negative energy state gets
filled whereby the Dirac sea is formed
\cite{dirac}\footnote{See for example \cite{weinberg} for
historical account.}.
This method of filling at first empty Dirac sea seems to make sense
only for fermions for which there is Pauli principle.
In this way ``correct vacuum'' is formed out of ``naive vacuum'', the
former well functioning phenomenologically.
Formally by filling Dirac sea we define creation operators
$b^+(\stackrel{\rightharpoonup}{p},s,\omega)$
for holes which is equivalent to destruction operators
$a(-\stackrel{\rightharpoonup}{p},-s,-\omega)$
for negative energies $-\omega$ and altogether opposite quantum
numbers.
This formal rewriting can be used also for bosons, but we have never
heard the associated filling of the negative energy states.

As a matter of fact the truely new content and the main motivation of
the present paper is to present an idea as to how the 2nd quantized
field
theory in the \underline{boson} case looks analogous to the fermion
system before the Dirac sea is filled out.
Although this bosonic theory analogous to the empty Dirac sea for
fermions has the serious drawbacks: It has an indefinite ``Hilbert
space'' as its Fock space.
Furthermore it possesses no bottom in the spectrum of the Hamiltonian.
However it has much nicer features than the true vacuum theory in which
the negative energy states are completely filled: Existence of position
eigenstates and description in terms of finite dimensional wave
functions.

At the very end when the true vacuum for the case of bosons is
realized according to our method presented in this paper, we will come
exactly the same theory as the usual one. Thus our approach cannot be
incorrect, but the true vacuum theory itself may not provide new results.

However ``the naively quantized theory'', which is an analogue of the
unfilled Dirac sea for fermions, is nice to think about because it turns out to be
a world in which only a few particles can be described by wave
functions of the positions of these few particles.
Remarkably, contrary to usual relativistic theories, the particles in
the ``naive vaccum world'' have position eigenstates.
They can be achieved only as superposition of positive and negative
energy eigenstates.

The problem of passage from the naive vacuum world to the usual theory
involves, as already mentioned, addition of ``minus one boson'' to each
negative energy state.
In the following section 2, we shall concretize how such idea of a
negative number of bosons can be thought upon mathematically by
treating the harmonic oscillator which is brought in
correspondence with a single particle
state under the usual second quantization.
We make an extension of the spectrum with the excitation number
$n=0,1,2,\ldots$ to the one with negative integer values
$n=-1,-2.\ldots$.
This extension can be performed by requiring that the wave function
$\psi(x)$ should be analytic in the whole complex $x$ plane except for
an essential singularity at $x=\infty$.
This requirement is a replacement to the usual condition on the norm
of the finite Hilbert space
$\int^\infty_{-\infty}|\psi(x)|^2d{\rm x}<\infty$.
The outcome of the study is that the harmonic oscillator has the
following two sectors :
1) the usual positive sector with positive and zero number of
particles, and
2) the negative sector with the negative number of particles.
The latter sector has indefinite Hilbert product.

But we would like to stress that there is a barrier between the usual
positive sector and the negative sector.
Due to the barrier it is impossible to pass from one sector to the other
with usual polynomial interactions.
This is due to some extrapolation of the
wellknown laser effect, which make easy to fill an already highly
filled single particle state for bosons.
This laser effect may become zero when an interaction tries to have
the number of particles pass the barrier.
In this way we may explain that the barrier prevents us from observing
a negative number of bosons.


It may be possible to use as a playground a formal world in which
one has neither yet filled the usual Dirac sea of fermions nor
performed the one boson removal from the negative energy state.
We shall indeed study such a playground model referred to as the naive
vacuum model.
Particularly we shall provide an analogous theorem to the CPT
theorem
\footnote{The CPT theorem is well explained in \cite{sakurai}},
since the naive vacuum is \underline{not} CPT invariant for
both fermions and bosons.
At first one might think that a strong reflection without associated
inversion of operator order might be good enough.
But it turns out this has the unwanted feature that the sign of
the interaction energy is not changed.
This changing the sign is required since under strong reflection the
sign of all energies should be switched.
To overcome this problem we propose the CPT-like symmetry for the
naive vacuum world to include further a certain analytic continuation.
This is constructed by applying a certain analytic continuation around
branch points which appear in the wave function for each
pair of particles.
It is presupposed that we can restrict our attention to such a family
of wave function as the one with sufficiently good physical
properties. The argument and proof of CPT-like theorem is deferred to
the successive paper[1].

We put forward a physical picture that may be of value in
developing an intuition on naive vacuum world.
In fact investigation of naive vacuum world may be very attractive
because the physics there is quantum mechanics of finite number of
particles.
Furthermore the theory is piecewise free in the sense that
relativistic interactions become of infinitely short range.
Thus the support that there are interactions is null set and one may
say that the theory is free almost everywhere.
But the very local interactions make themselves felt only via boundary
conditions where two or more particles meet.
This makes the naive vacuum world a theoretical playground.
However it suffers from the following severe drawbacks from a
physical point of view :

\begin{itemize}
\item No bottom in the Hamiltonian
\item Negative norm square states
\item Pairs of particles with tachyonically moving center of mass
\item It is natural to work with ``anti-bound states'' rather that
bound states in the negative energy regime.
\end{itemize}

What we really want to present in the present article is a more
dramatic formulation of relativistic second quantization
of boson theory
and one may think of it as a
quantization procedure.
We shall formulate below the shift of vacuum for bosons as a shift of
boundary conditions in the wave functional formulation of the second
quantized theory.

But, using the understanding
of second quantization of particles along
the way we describe, could we get a better understanding as to how to
second quantize strings?
This is our original motivation of the present work.
In the oldest attempt to make string field theory by Kaku and
Kikkawa \cite{kaku} the infinite momentum frame was used.
To us it looks like an attempt to escape the problem of the negative
energy states.
But this is the root of the trouble to be resolved by the modification
of the vacuum described above.
So the hope would be that by grasping better these Dirac sea problems
in our way, one might get the possibility of inventing new types of
string field theories, where the infinite momentum frame would not
be called for.

The present paper is organized as follows.
Before going to the real description of how to quantize bosons in our
formulation we shall formally look at the
harmonic oscillator in the following section 2.
It is naturally extended to describe a single particle state that can
also have a negative number of particles in it.
In section 3 application to even spin particles is described, where
the negative norm square problems are gotten rid of.
In section 4 we bring our method into a wave functional formulation,
wherein changing the convergence and finite norm
conditions are explained.
In section 5 we illustrate the main point of the formulation of the
wave functional by considering a double harmonic oscillator.
This is much like a $0+1$ dimensional world instead of the usual $3+1$
dimensional one.
In section 6 we go into a study of the naive vacuum world.
Finally in section 7 
we give conclusions.


\section{The analytic harmonic oscillator}

In this section we consider as an exercise the formal problem of the
harmonic oscillator with the requirement of analyticity of the wave
function.
This will turn out to be crucial for our treatment of bosons with a
Dirac sea method analogous to the fermions.
In this exercise the usual requirement that the wave function
$\psi(x)$ should be square integrable

\begin{equation}
\int^\infty_{-\infty}|\psi(x)|^2dx<\infty
\end{equation}
is replace by the one that
\begin{equation}
\psi(x) {\rm \ is \ analytic \ in \ C}
\end{equation}
where a possible essential singularity at $x=\infty$ is allowed.
In fact for this harmonic oscillator we shall prove the following
theorem :

1) The eigenvalue spectrum $E$ for the equation
\begin{equation}
\left(-\frac{\hbar^2}{2m}\frac{\partial^2}{\partial x^2}+
\frac{1}{2}m\omega^2x^2\right)\psi(x)=E\psi(x)
\end{equation}
is given by
\begin{equation}
E=(n+\frac{1}{2})\hbar\omega \ \ \\ (n\varepsilon Z)
\end{equation}
with {\it any integer} $n$.

2)The wave functions for $n=0,1,2,\ldots$ are the usual ones

\begin{equation}
\varphi_n(x)=A_n e^{-\frac{1}{2}(\beta x)^2}H_n(\beta x) \ \ \ .
\end{equation}

Here $\beta^2=\frac{m\omega}{\hbar}$ and $H_n(\beta x)$ the Hermite
polynomials of $\beta x$ while
$A_n=\sqrt{\frac{\beta}{\pi^{\frac{1}{2}}2^nn!}}$ .
For $n=-1,-2,\ldots$ the eigenfunction is given by

\begin{equation}
\varphi_n(x)=\varphi_{-n-1}(ix)=A_{-n-1}e^{\frac{1}{2}(\beta x)^2}
	H_{-n-1}(i\beta x) \ \ \ .
\end{equation}

3)The inner product is defined as the natural one given by

\begin{equation}
<\psi_1|\psi_2>=\int_\Gamma\psi_1(x^*)^*\psi_2(x)dx
\label{2.7}
\end{equation}
where the contour denoted by $\Gamma$ is taken to be the one along the
real axis from $x=-\infty$ to $x=\infty$.
The $\Gamma$ should be chosen so that the integrand should go down
to zero at $x=\infty$, but there remains some
ambiguity in the choice of  $\Gamma$.
However if one chooses the same $\Gamma$
for
 all the negative $n$
states, the norm squares of these states have an alternating sign.
In fact for the path $\Gamma$ along the imaginary axis from
$-i\infty$ to $i\infty$, we obtain

\begin{eqnarray}
<\varphi_n|\varphi_m>&=&
	\int^{i\infty}_{-i\infty}\varphi_n(x^*)^*\varphi_n(x)dx\nonumber\\
&=&-(-1)^n
\label{2.8}
\end{eqnarray}

The above 1)--3) constitute the theorem.

Proof of this theorem is rather trivial.
We may start with consideration of large numerical $x$ behavior of a
solution to the eigenvalue equation.
Ansatze for the wave function is made in the form

\begin{equation}
\psi(x)=f(x)e^{\pm\frac{1}{2}(\beta x)^2}
\end{equation}
and we rewrite the eigenvalue equation(2.3) as

\begin{equation}
\frac{f''(x)}{\beta^2f(x)}\pm\frac{2f'(x)}{\beta f(x)}\beta x =
	-\frac{E\mp\frac{1}{2}\omega\hbar}{\omega\hbar}\ \ \ .
\label{2.10}
\end{equation}
If we use the approximation that the term $f''(x)/\beta^2f(x)$ is
dominated by the term $\pm\frac{2f'(x)}{\beta f(x)}\beta x$ for large
$|x|$, eq.(\ref{2.10}) reads

\begin{equation}
\frac{d\log f(x)}{d\log
x}=\frac{\mp E+\frac{1}{2}\omega\hbar}{\omega\hbar} \ \ \ .
\end{equation}
Here the right hand side is a constant $n$ which is yet to be shown to
be an integer and we get as the large $x$ behavior

\begin{equation}
f(x)\sim x^n
\end{equation}
The reason that $n$ must be integer is that the function $x^n$ will
otherwise have a cut.
Thus requiring that $f(x)$ be analytic except for $x=0$ we must have

\begin{equation}
\mp E=-\frac{1}{2}\omega\hbar + n\hbar\omega
\end{equation}
For the upper sign the replacement $n\rightarrow -n-1$ is made and we can
always write

\begin{equation}
E=\frac{1}{2}\hbar\omega+n\hbar\omega
\end{equation}
where $n$ takes not only the positive and zero integers
$n=0,1,2,\ldots$, but also the negative series $n=-1,-2 \ldots$ .

Indeed it is easily found that for negative $n$ the wave function is

\begin{equation}
\varphi_n(x)=\varphi_{-n-1}(ix) = A_{-n-1}e^{\frac{1}{2}(\beta
x)^2}H_{-n-1}(i\beta x)
\end{equation}

Next we go to the discussion of the inner product which we define by
eq.(\ref{2.7}).
If the integrand $\psi_1(x^*)^*\psi_2(x)$ goes to zero as
$x\rightarrow\pm\infty$ the contour $\Gamma$ can be deformed as
usual.
But when the integrand does not go to to zero, we may have to define inner
product by an analytic continuation of the wave functions from the
usual positive sector ones that satisfy
$\int|\psi(x)|^2dx<\infty$ .
If we choose $\Gamma$ to be the path along the imaginary axis from
$x=-i\infty$ to $x=i\infty$, the inner product takes the form

\begin{eqnarray}
<\varphi_n|\varphi_m>&=&
	\int^{i\infty}_{-i\infty}\varphi_n(x^*)^*\varphi_m(x)dx\nonumber\\
&=&i\int^\infty_{-\infty}
\varphi_{-n-1}\left(i\left(i\xi\right)^*\right)^*
	\varphi_{-m-1}\left(i\left(i\xi\right)\right)d\xi
\label{2.16}
\end{eqnarray}
where $x$ along the imaginary axis is parameterized by $x=i\xi$ with a
real $\xi$ .
From eq.(\ref{2.16}) we obtain for the negative $n$ and $m$,

\begin{equation}
<\varphi_n|\varphi_m>=-i(-1)^m\delta_{nm}
\label{2.17}
\end{equation}
so that

\begin{equation}
\parallel\varphi_n\parallel^2=-i(-1)^n \ \ \ .
\label{2.18}
\end{equation}
We notice that the norm square has the alternating sign, apart from a
prefactor $-i$, depending on
the even or odd negative $n$, when the contour $\Gamma$ is kept fixed.

The reason why there is a factor $i$ in eq.(\ref{2.18}) can be understood as
follows:
When the complex conjugation for the definition of the inner product
(\ref{2.7}) is taken, the contour $\Gamma$ should also be complex
conjugated

\begin{equation}
<\psi_1|\psi_2>^*=\int_{\Gamma^*}\psi_1(x^*)\psi_2(x)^* dx
\end{equation}
Thus if $\Gamma$ is described by $x=x(\xi)$ as

\begin{equation}
\Gamma=\{x(\xi)|-\infty<\xi<\infty \ \ \ : \ \ \ \xi ={\rm real}\}
\end{equation}
then $\Gamma^*$ is given by

\begin{equation}
\Gamma^*=\{x^*(\xi)|-\infty<\xi<\infty\ \ \ ,\ \ \ \xi={\rm real}\} 
 \ .
\end{equation}
So we find

\begin{equation}
<\psi_1|\psi_2>^*=
	\int_{-\infty<\xi<\infty}\psi_2(x(\xi)^*)^*\psi_1(x(\xi))
	\frac{dx(\xi)^*}{dx(\xi)} dx(\xi)
\end{equation}
which deviates from $<\psi_2|\psi_1>$ by the factor
$dx(\xi)^*/dx(\xi)$ in the integrand.
In the case of $x(\xi)=i\xi \ , \ dx(\xi)^*/dx(\xi)=-1$ so that

\begin{equation}
<\psi_1|\psi_2>^*=-<\psi_2|\psi_1>
\end{equation}
for the eigenfunctions of the negative sector.
From this relation the norm square is purely imaginary.

This convention of the inner product may be strange one and  we may
change the inner product eq.(\ref{2.7}) by a new one defined by

\begin{equation}
<\psi_1|\psi_2>_{new}=\frac{1}{i}<\psi_1|\psi_2>
\end{equation}
so as to have the usual relation also in the negative sector

\begin{equation}
<\psi_1|\psi_2>^*_{new}=<\psi_2|\psi_1>_{new}
\end{equation}
if we wish.


\section{The treatment of the Dirac sea for bosons}

In this section we shall make use of the extended harmonic oscillator described in
previous section to quantize bosons.

As is well known in a non-relativistic theory a second quantized
system of bosons may be described by using an analogy with a system
of harmonic oscillators ;
one for each state in an orthonormal basis for the single particle.
The excitation number $n$ of the harmonic oscillator is identified
with the number of bosons present in that state in the basis to which
the oscillator corresponds.

For instance, if we have a system with $N$ bosons its state is
represented by the symmetrized wave function

\begin{equation}
\psi_{\alpha_1\ldots\alpha_N} \ ( \stackrel{\rightharpoonup}{x_1},
\ldots , \stackrel{\rightharpoonup}{x_N})
\end{equation}
where the indices $\alpha_1,\alpha_2\ldots,\alpha_N$ indicate the
intrinsic quantum numbers such as spin.
In a energy and momentum eigenstate
$k=(\stackrel{\rightharpoonup}{k},+)$ or
$k=(\stackrel{\rightharpoonup}{k},-)$
where the signs $+$ and $-$ denote those of the energy, 
we may write

\begin{eqnarray}
K_{\rm pos}&=&\{(\stackrel{\rightharpoonup}{k},+) |
\stackrel{\rightharpoonup}{k}\} \ , \\
K_{\rm neg}&=&\{(\stackrel{\rightharpoonup}{k},-) |
\stackrel{\rightharpoonup}{k}\} \  .
\end{eqnarray}
and $K=K_{\rm pos}\cup K_{\rm neg}$ .
We expand
$\psi_{\alpha_1\ldots\alpha_N}
(\stackrel{\rightharpoonup}{x_1},\ldots,\stackrel{\rightharpoonup}{x_N})$
in terms of an orthonormal basis of single particle states
$\{\varphi_{k;\alpha} \ (\stackrel{\rightharpoonup}{x})\}$
with
$k\epsilon K$.
It reads

\begin{eqnarray}
|\psi>&=&\psi_{\alpha_1\ldots,\alpha_N}(\stackrel{\rightharpoonup}{x_1},
	\ldots,\stackrel{\rightharpoonup}{x_N})\nonumber\\
&=&\sum C_{{k_1},\ldots,k_N}\frac{1}{N!}\sum_{\rho\epsilon S_N}\nonumber\\
&&\varphi_{k_{\rho(1)}\alpha_1}
		(\stackrel{\rightharpoonup}{x_1})
	\varphi_{k_{\rho(2)}\alpha_2}
		(\stackrel{\rightharpoonup}{x_2})\cdots
	\varphi_{k_{\rho(N)}\alpha_N}
		(\stackrel{\rightharpoonup}{x_N})\ \ \ .
\end{eqnarray}
The corresponding state of the system of the harmonic oscillators is given
by

\begin{equation}
|\psi > = \sum_{k_1,\ldots,k_N} C_{k_1,\ldots k_N}
	\prod_{k\epsilon K}
| n_k >
\end{equation}
where $|n_k>$ represents the state of the $k$-th harmonic oscillator.

The harmonic oscillator is extended so as to have the negative $n_k$
values of the excitation number.
This corresponds to that the number of bosons $n_K$ in the single particle 
states could be negative.
In the non-relativistic case one can introduce the creation and
annihilation operators $a_k$ and $a^+_k$ respectively.
In the harmonic oscillator formalism these are the step operators for
the $k$th harmonic oscillator,

\begin{eqnarray}
a^+_k |n_k> &=&\sqrt{n_k+1} \ | n_k+1>\\
a_k|n_k > &=& \sqrt{n_k} \ | n_k-1> \ \ .
\end{eqnarray}

It is also possible to introduce creation and annihilation operators
for arbitrary states $|\psi > $

\begin{eqnarray}
a^+(\psi)&=&\sum_{k\epsilon K}<\varphi_k|\psi> a^+_k\\
a(\psi)&=&\sum_{k\epsilon K}a_k<\varphi_k|\psi> \ \ ,
\end{eqnarray}

\noindent
where the inner product is defined by $\int d^3x\varphi^*(x)i\stackrel{\leftrightarrow}{\partial}_0\psi(x)$.

We then find

\begin{eqnarray}
[a(\psi'),a^+(\psi)]&=&\sum_{k,k'}<\psi'|\varphi_{k'}>
	[a_{k'},a_k]<\varphi_k|\psi>\nonumber\\
&=&<\psi'|\psi> \ \ .
\end{eqnarray}
in which the right hand side contains an indefinite Hilbert product.
Thus if we perform this naive second quantization, the possible
negative norm square will be inherited into the second quantized
states in the Fock space.

Suppose that we choose the basis such that for some subset
$K_{\rm pos}$ the norm square is unity

\begin{equation}
<\varphi_k|\varphi_k>= 1 \ \ \  {\rm for} \ \  k\epsilon K_{\rm pos}
\end{equation}
while for the complement set
$K_{\rm neg}=K\backslash K_{\rm pos}$ it is

\begin{equation}
<\varphi_k|\varphi_k>=-1\ \ \ {\rm for} \ k\epsilon K_{\rm neg} \ \ .
\end{equation}
Thus any component of a Fock space state must have negative norm
square if it has an odd number of particles in states of
$K_{\rm neg}$.

We thus have the following
signs of the norm square in the naive second quantization

\begin{equation}
<n_k|m_k> = \delta_{n_km_k}(-1)^{n_k}
\end{equation}
for $k\epsilon K_{\rm neg}$ where $n_k$ and $m_k$ denote the usual
nonzero levels.
With use of our extended harmonic oscillators we end up with a system
of norm squared as follows :

\noindent
For $k\epsilon K_{\rm pos}$

\begin{equation}
\begin{array}{ll}
&<n_1,\ n_2\ldots|m_1,\ m_2, \ldots> \\
&=\left\{
\begin{array}{ll}
\delta_{n_km_k}&{\rm for} \ n_k,m_k=0,1,2,\ldots\\
i\delta_{nkm_k}(-1)^{n_k}& {\rm for} \ n_k,m_k=-1,-2\ldots\\
\infty&{\rm for} \ n_k \ {\rm and} \ m_k \ {\rm in \ different \ sectors}.
\end{array}
\right.
\end{array}
\end{equation}

\noindent
For $k\epsilon K_{\rm neg}$
\begin{equation}
\begin{array}{ll}
&<n_1,\ n_2\ldots|m_1,\ m_2, \ldots> \\
&=\left\{
\begin{array}{ll}
\delta_{n_km_k}(-1)^{n_k}&{\rm for} \ n_k,m_k=0,1,2,\ldots\\
i\delta_{n_km_k}& {\rm for} \ n_k,m_k=-1,-2\ldots\\
\infty&{\rm for} \ n_k \ {\rm and} \ m_k \ {\rm in \ different \ sectors} .
\end{array}
\right.
\end{array}
\end{equation}
We should bear in mind that the trouble of negative norm square can be
solved by putting {\it minus one particle} into each state with
$k\epsilon K_{\rm neg}$.
Thereby we get it restricted to negative number of particles in these
states.
Thus we have to use the inner product
$<n_k|m_k>=i\delta_{n_km_k}$, which makes the Fock space sector be a
good positive definite Hilbert space apart from the overall factor
$i$.

We may formulate our procedure in the following.
The naive vacuum may be described by the state in terms of the ones of
the harmonic oscillators as

\begin{equation}
| \,{\rm naive \ vac.} > =
	 \prod_{k\epsilon K} | 0>_{\stackrel{kth}{osc}} \ \ \ .
\end{equation}
where $|0>_{\stackrel{kth}{osc}}$ denotes the vacuum state of the $kth$
harmonic oscillator.

On the other hand the correct vacuum is given by the state

\begin{equation}
| \,{\rm correct \ vac.} > =
	\prod_{k\epsilon K_{\rm pos}}|0>_{\stackrel{kth}{osc}} \cdot
	\prod_{k\epsilon K_{\rm neg}} | -1 >_{\stackrel{kth}{osc}}
\end{equation}
where the states $|-1>$ in $K_{\rm neg}$ are the ones with minus one
particles.

We may proceed to the case of relativistic integer spin particles of which
inner product is indefinite by Lorentz invariance

\begin{equation}
\int\psi^\ast(\stackrel{\rightharpoonup}{x},t)
	\stackrel{\leftrightarrow}{\partial_t}
	\psi(\stackrel{\rightharpoonup}{x},t) d^3
	\stackrel{\rightharpoonup}{x} \ \ \ .
\end{equation}

For the simplest scalar field case, the energy of the naive vacuum is given by

\begin{equation}
{\rm E_{naive \ vac.}}=
	\sum_{k\epsilon K}\frac{1}{2}\omega_k = 0 \ \ \ .
\end{equation}
By adding minus one particle to each negative energy state
$\varphi_{k-}$ with $k\epsilon K_{\rm neg}$ the second quantized system
is brought into such a sector that it is in the ground state, which is
the correct vacuum.
The energy of it is given by

\begin{eqnarray}
{\rm E_{correct \ vac.}}&=&
	\sum_{k\epsilon K}\frac{1}{2}\omega_k-
	\sum_{k\epsilon K_{\rm neg}}\frac{1}{2}\omega_k\\
&=&\sum_{k\epsilon K}\frac{1}{2}|\omega_k | =
	\sum_{k\epsilon K_{\rm pos}}\omega_k \ \ \ .
\end{eqnarray}
It should be stressed that only inside the sector we obtain the ground
state in this way.
In fact with the single particle negative energies for bosons, the
total hamiltonian may have no bottom.
So if we do not add minus one particle to each single particle negative
energy state, one may find a series of states of which energy goes to
$-\infty$.
However, by adding minus one particle we get a state of the second
quantized system in which there is the barrier due to the laser
effect.
This barrier keeps the system from falling back to lower energies as
long as polynomial interaction in $a^+_k$ and $a_k$ are concerned.

In the above calculation for the relativistic case we have

\begin{equation}
{\rm E_{correct \ vac}> E_{naive \ vac}} \ \ \ .
\end{equation}
Thus at the first sight the correct vacuum looks unstable.
However which vacuum has lower energy is not important for the
stability of a certain proposal of vacuum.
Rather the range of allowed energies for the sector of the vacuum
proposal is important.
To this end we define the energy range ${\rm E_{range}}$ of the vacuum
by

\begin{equation}
{\rm E_{range}}(|{\rm vac}>)=\{{\rm E}\}
\end{equation}
where ${\rm E}$ denotes an energy in a state which can be reached from
$|{\rm vac}>$ by some operators polynomial in $a^+$ and $a$.
Thus for the naive vacuum

\begin{equation}
{\rm E_{range}(|naive \ vac>)}=(-\infty,\infty)
\end{equation}
while for the correct vacuum

\begin{equation}
{\rm E_{range}(|correct \ vac>)}=
	[\sum_{k\epsilon K}\frac{1}{2}|\omega_K|, \infty] \ \ \ .
\end{equation}

Once the vacuum is brought into the correct vacuum state, it is no
longer possible to add particles to the state with $K_{neg}$, due to
the barrier: It is  rather to subtract particles.
Thus $a_k$ with $k\epsilon K_{\rm neg}$ can act on
$|-1>_{\stackrel{kth}{osc}}$ with arbitrary number of times as

\begin{equation}
(a_k)^n |-1>_{\stackrel{kth}{osc}}=
	\sqrt{|n|!} \ | -1-n >_{\stackrel{kth}{osc}} \ \ \ .
\end{equation}
These subtractions we may call holes which correspond to addition of
antiparticles.

It is natural to switch notations from dagger to non dagger one by
defining

\begin{equation}
b^+(-\stackrel{\rightharpoonup}{k},{\rm anti})=
	a(\stackrel{\rightharpoonup}{k}, -)
\end{equation}
and vice versa
where $k=(\stackrel{\rightharpoonup}{k}, -)$ is a $\omega<0$ state
with 3-momentum $\stackrel{\rightharpoonup}{k}$.
The operator $b^+(-\stackrel{\rightharpoonup}{k}, {\rm anti})$
denotes a creation of antiparticle with momentum
$\stackrel{\rightharpoonup}{k}$ and positive energy $-\omega>0$ .
This is exactly the usual way of treatment of the second quantization
for bosons.
The commutator of these operators reads
\begin{equation}
[b(\stackrel{\rightharpoonup}{k}, {\rm anti}),
	b^+(\stackrel{\rightharpoonup}{k'},{\rm anti})]=
	\delta_{\stackrel{\rightharpoonup}{k}
		\stackrel{\rightharpoonup}{k'}} \ \ \ .
\end{equation}
It should be noticed that in the boson case the antiparticles are also
holes.
Before closing this section two important issues are discussed.
The first issue is that there are potentially possible four vacua in
our approach of quantization.

We have argued that we can obtain the correct vacuum by modifying the
naive vacuum so that one fermion is filled and one boson removed from
each single particle negative energy state.
This opens the possibility of considering naive vacuum and associated
world of states where there exist a few extra particles.
The naive vacuum should be considered as a playground for study of the
correct vacuum.
It should be mentioned that once we start with one of the vacua and
work by filling the negative energy states or removing from it, we may
also do so for positive energy states.
In this way we can think of four different vacua which are
illustrated symbolically as type a-d in Fig.1.

\begin{figure}[h]
\epsfysize=120mm
\epsfbox{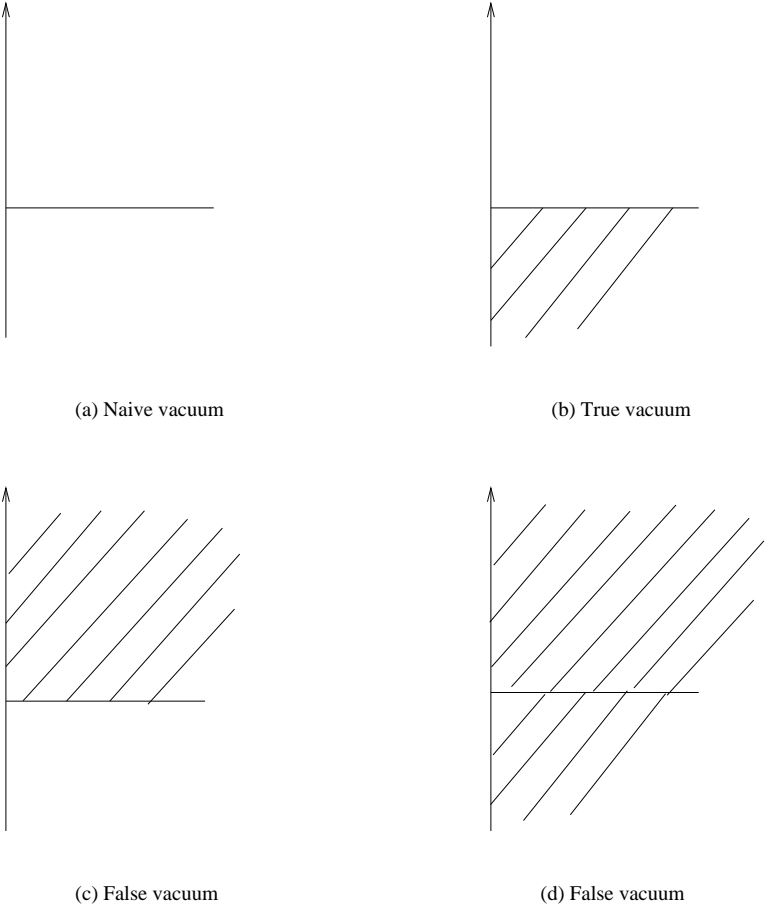}
\caption{Four types of vacua}{
There are four possible types of vaua for bosons as well as fermions.
Here the vertical axis indicates energy level. In Figures (a) - (d) the
shaded states denote that they are all filled
by one particle for fermions and minus one particle for bosons.
The unshaded states are empty.
}
\end{figure}

As an example let us consider the type (c) vacuum.
In this vacuum the positive energy states are modified by filling the
positive energy states by one fermion but removing from it by one
boson, while the negative energy states are not modified.
Thus the single particle energy spectrum has a top but no bottom.
Inversion of the convention for the energy would not be possible to
be distinguished by experiment as far as free system is concerned.
However it would have negative norm square for all bosons and the
interactions would work in an opposite manner.
We shall show in the later sections that there exists a trick of
analytic continuation of the wave function to circumvent this
inversion of the interaction.

Another issue to be mentioned is the CPT operation on those four
vacua.
The CPT operation on the naive vacuum depicted as type (a) vacuum in
Fig.1 does not get it back.
The reason is that by the charge conjugation operator C all the holes
in the negative energy states are, from the correct vacuum point
of view, replaced by particles of corresponding positive energy
states.
Thus acting CPT operator on the naive vacuum is sent into the type (c)
vacuum because the positive energy states is modified while the
negative ones remain the same.
This fact may be stated that in the naive vacuum CPT symmetry is
spontaneously broken.

However in the subsequent paper ``Dirac sea for bosons II''[1] we shall
put forward another CPT-like theorem in which the CPT-like symmetry is
preserved in the naive vacuum but broken in the correct one.

Before closing this section, 
we mention some properties of the world around the naive vacuum
where there is only a few particles.
The terminology of ``the world around a vacuum'' is used for
Hilbert space
with
a superposition of such a states that it deviates from the vacuum in
question by a finite number of particles and that the boson does not
cross the barrier.
Since the naive vacuum has no particle and we can add positive number
of particles which, however, can have both positive and negative
energies.
The correct vacuum may similarly have a finite number of particles and
holes in addition to the negative energy seas.

\section{Wave functional formulation}

In this section we develop the wave functional formulation of field
theory in the naive vacuum world.

When going over field theory in the naive field
quantization

\begin{eqnarray}
\varphi(\stackrel{\rightharpoonup}{x},t)&=&
	\sum_{\stackrel{\rightharpoonup}{p},{\rm sig}n}
	\frac{1}{\sqrt{|\omega |}}
	a(\stackrel{\rightharpoonup}{p},{\rm sig}n)
	e^{-i\omega t+i\stackrel{\rightharpoonup}{p}\cdot
		\stackrel{\rightharpoonup}{x}}
\\
\pi(\stackrel{\rightharpoonup}{x},t)&=&
	\sum_{\stackrel{\rightharpoonup}{p},{\rm sig}n}
	\frac{1}{\sqrt{|\omega |}}
	a(\stackrel{\rightharpoonup}{p},{\rm sig}n)
	\cdot({\rm sig}n)=
	e^{-i\omega t+i\stackrel{\rightharpoonup}{p}\cdot
		\stackrel{\rightharpoonup}{x}}
\end{eqnarray}
we have a wave functional $\Psi[\varphi]$.
For each eigenmode
$\omega\varphi_{\stackrel{\rightharpoonup}{p}}+
	i\pi_{\stackrel{\rightharpoonup}{p}}$ ,
where $\varphi_{\stackrel{\rightharpoonup}{p}}$ is the 3-spatial
Fourier transform of $\varphi(x)$ and
$\pi_{\stackrel{\rightharpoonup}{p}}$ is conjugate momentum, we have
an extended harmonic oscillator described in section 2.
In order to see how to put the naive vacuum world into a wave
functional formulation, we investigate the Hamiltonian and the
boundary conditions for a single particle states with a general norm
square.

Let us imagine that we make the convention in which the $n$-particle
state be

\begin{equation}
A_nH_n(x)
\end{equation}
with $H_n$ the Hermite polynomial.
Thus

\begin{equation}
|n>=A_nH_n(x)|0> \ \ \ .
\end{equation}
On the other hand $n$ excited state in the harmonic oscillator is given
by

\begin{equation}
A_nH_n(x)\beta e^{-\frac{1}{2}(\beta x)^2}
\end{equation}
with $\beta^2=\frac{\hbar}{m\omega}$.
We can vary the normalization while keeping the convention
\begin{equation}
<n|n>=\beta^{-2n}<0|0> \ \ \ .
\end{equation}
We may consider $\beta^{-2}$ as the norm square of the single particle
state corresponding to the harmonic oscillator.

Now the Hamiltonian of the harmonic oscillator is expressed in terms
of
$\omega$ and $\beta^{-2}$ as follows:

\begin{equation}
H=-\frac{\omega}{2<s.p|s.p>}\frac{d^2}{dx^2}+\frac{1}{2}<s.p|s.p>
\omega x^2
\label{4.7}
\end{equation}
where $|s.p>$ denotes the single particle state and thus

\begin{equation}
<s.p|s.p>=m\omega=\beta^{-2}
\end{equation}
with $\hbar=1$. Therefore we obtain the Hamiltonian

\begin{equation}
H=-\frac{1}{2}\beta^2\omega\frac{d^2}{dx^2}+\frac{1}{2}\beta^{-2}\omega x^2
\end{equation}

Remark that if one wants $<s.p|s.p>$ negative for negative $\omega$
one, $\beta$ turns out to be pure imaginary. Thus
$e^{-\frac{1}{2}(\beta \chi)^2}$ blows up so that the wave functions
become like the one in the extended negative sector discussed in the
previous sections.

By passing to the correct vacuum world by removing one particle from
each negative energy state, the boundary conditions for the wave
functional are changed so as to converge along the real axis for all
the modes.
Remember they are along the imaginary axis for the negative energy
modes in the naive vacuum.

From the fact that the form of the Hamiltonian in the wave functional
formalism must be the same as for the correct vacuum we can easily
write down the Hamiltonian.
For instance using the conjugate variable $\pi$

\begin{equation}
\pi(\stackrel{\rightharpoonup}{x})=-i
	\frac{\delta}{\delta\varphi(\stackrel{\rightharpoonup}{x})}
\end{equation}
the free Hamiltonian becomes

\begin{equation}
H_{\rm free}=\int\frac{1}{2}\left\{|\pi(\stackrel{\rightharpoonup}{x})|^2+
	|\bigtriangledown\varphi(\stackrel{\rightharpoonup}{x})|^2+
	m|\varphi(\stackrel{\rightharpoonup}{x})|^2\right\}
	d^3\stackrel{\rightharpoonup}{x} \ \ \ .
\end{equation}
This acts on the wave functional as

\begin{eqnarray}
&&H_{\rm free}\Psi[\varphi]\nonumber\\
&&=\frac{1}{2}\int\left\{-
	\frac{\delta^2}{\delta\varphi(\stackrel{\rightharpoonup}{x})^2}+
	|\bigtriangledown\varphi(\stackrel{\rightharpoonup}{x})|^2+
	m^2|\varphi(\stackrel{\rightharpoonup}{x})|^2\right\}
	\Psi[\varphi]\ \ \ .
\end{eqnarray}
The inner product for the functional integral is given by

\begin{eqnarray}
<\Psi_1|\Psi_2>&=&
	\int\Psi_1[(Re\varphi)^\ast,
		(Im\varphi)^\ast]^\ast\nonumber\\
&&\cdot\Psi_2[Re\varphi,Im\varphi]
	{\cal D}Re\varphi\cdot{\cal D}Im\varphi \ \ \ ,\nonumber
\end{eqnarray}
where the independent functions are
$Re\varphi(\stackrel{\rightharpoonup}{x})$ and
$Im\varphi(\stackrel{\rightharpoonup}{x})$ .
In order to describe the wave functional theory of the naive vacuum
world we shall make a formulation in terms of the convergence
condition along the real function space for
$Re\varphi$ and $Im\varphi$ .
In fact we go to the representation in which
$\Psi [Re\varphi, Im\pi]$ is expressed by means of $\varphi$ and
$\pi$.

We would like to organize so that the boundary conditions for the
quantity
$\omega\varphi_{\stackrel{\rightharpoonup}{k}}+
i\pi_{\stackrel{\rightharpoonup}{k}}$
are convergent in the real axis for $\omega>0$ while for $\omega<0$ 
they are so in the imaginary axis.
We may think the real and imaginary parts of
$(\omega\varphi_{\stackrel{\rightharpoonup}{k}}+
i\pi_{\stackrel{\rightharpoonup}{k}})$
separately.
Then the requirement of convergence in the correct vacuum should be
that for $\omega<0$ the formal expression

\begin{eqnarray}
Re(\omega\varphi_{\stackrel{\rightharpoonup}{k}}+
i\pi_{\stackrel{\rightharpoonup}{k}})&=&
	\frac{\omega}{2}\left\{(Re\varphi)_{\stackrel{\rightharpoonup}{k}}+
		(Re\varphi)_{-\stackrel{\rightharpoonup}{k}}\right\}
		\nonumber\\
&&-\frac{1}{2}\left\{(Im\pi)_{\stackrel{\rightharpoonup}{k}}+
		(Im\pi)_{-\stackrel{\rightharpoonup}{k}}\right\}
\end{eqnarray}	
and

\begin{eqnarray}
Im(\omega\varphi_{\stackrel{\rightharpoonup}{k}}+
i\pi_{\stackrel{\rightharpoonup}{k}})&=&
	\frac{\omega}{2}\left\{(Im\varphi)_{\stackrel{\rightharpoonup}{k}}+
		(Im\varphi)_{-\stackrel{\rightharpoonup}{k}}\right\}
		\nonumber\\
&&+\frac{1}{2}\left\{(Re\pi)_{\stackrel{\rightharpoonup}{k}}+
		(Re\pi)_{-\stackrel{\rightharpoonup}{k}}\right\}
\end{eqnarray}
are purely imaginary along the integration path for which the
convergence is required.

We may use the following parameterization in terms of the two real
functions $\chi_1$ and $\chi_2$ :

\begin{eqnarray}
Re\varphi&=&-(1+i)\chi_1-(1-i)\chi_2\nonumber\\
Im\pi&=&(1-i)\chi_1+(1+i)\chi_2\nonumber
\end{eqnarray}
By this parameterization the phases of
$\omega\varphi_{\stackrel{\rightharpoonup}{k}}+
	i\pi_{\stackrel{\rightharpoonup}{k}}$
lay in the intervals

\begin{eqnarray}
&&]-\frac{\pi}{4} \ , \ \frac{\pi}{4}[ \ \ \
	{\rm for} \ \ \  \omega>0\nonumber\\
{\rm and }&&\nonumber\\
&&]\frac{\pi}{4} \ , \ \frac{3\pi}{4}[ \ \ \
	{\rm for} \ \ \  \omega<0\nonumber
\end{eqnarray}
modulo $\pi$.
They provide the boundary conditions for the naive vacuum world when
convergence of ${\cal D}\chi_1{\cal D}\chi_2$ integration is required.

In this way  we find the naive vacuum world with usual wave functional
hamiltonian operator.
However, we do not require the usual convergence condition

\begin{equation}
\int\Psi(Re\varphi, Im\varphi)^\ast
\Psi(Re\varphi, Im\varphi){\cal D}Re\varphi{\cal D}Im\varphi < \infty
\label{4.14}
\end{equation}
but instead require

\begin{equation}
<\Psi|\Psi>=\int\Psi[(Re\varphi)^\ast,(Im\pi)^\ast]^\ast
	\Psi[Re\varphi, Im\varphi]{\cal D}\chi_1{\cal D}\chi_2 <
\infty
\end{equation}
where the left hand side is defined along the path with
$\chi$-parameterization.
The inner product corresponding to this functional contour is

\begin{eqnarray}
&<\Psi_1|\Psi_2>&=
	\int\Psi_1[-(1-i)\chi_1-(1+i)\chi_2,(1+i)\chi_1+(1-i)\chi_2]^\ast
	\nonumber\\
&&\Psi_2[-(1+i)\chi_1-(1-i)\chi_2,(1-i)\chi_1+(1+i)\chi_2]
	{\cal D}\chi_1{\cal D}\chi_2 \ \ \ .
\label{4.16}
\end{eqnarray}
This is not positive definite, and that is related to the fact that there
are lot of negative norm square states in the Fock space in the naive
vacuum world.

The method of filling the Dirac sea vacuum for fermions is now
extended to the case
of bosons that first in the naive vacuum we have the strange convergence
condition eq.(\ref{4.14}).
We then go to the correct vacuum by switching the boundary conditions
to the convergence ones along the real axis e.g. $Re\varphi$ and $Im\pi$ real.


\section{Double harmonic oscillator}

To illustrate how our functional formalism works we consider as a simple
example a double harmonic oscillator.
It is relevant for us in the following three reasons:

\begin{enumerate}
\renewcommand{\labelenumi}{\arabic{enumi})}
\item It is the subsystem of field theory which consists of two single
particle states with
$p^\mu=
(\stackrel{\rightharpoonup}{p},\omega(\stackrel{\rightharpoonup}{p}))$
and
$-p^\mu=
(-\stackrel{\rightharpoonup}{p},-\omega(\stackrel{\rightharpoonup}{p}))$
for $\omega(\stackrel{\rightharpoonup}{p})>0$ .
\item It could correspond a single 3-position field where
the gradient interaction is ignored.
\item It is a $0+1$ dimensional field theory model.
\end{enumerate}

We start by describing the spectrum for free case corresponding to a
two state system in which
the two states have opposite $\omega$'s.
The boundary conditions in the naive vacuum world is given by

\begin{eqnarray}
&&\int
\psi\left(\left(Re\varphi\right)^\ast , \left(Im\Pi\right)^\ast\right)^\ast
\psi\left(Re\varphi , Im\Pi\right)d\chi_1\chi_2
\nonumber\\
&&=\int
\psi\left(-\left(1-i\right)\chi_1-\left(1+i\right)\chi_2 ,
	\left(1+i\right)\chi_1+\left(1-i\right)\chi_2\right)^\ast
	\nonumber\\
&&\cdot \ \psi\left(-\left(1+i\right)\chi_1-\left(1-i\right)\chi_2 ,
	\left(1-i\right)\chi_1+\left(1+i\right)\chi_2\right)
	d\chi_1d\chi_2 < \infty
\label{5.1}
\end{eqnarray}
which is similar to eq.(\ref{4.16}).
However in eq.(\ref{5.1})
the quantities $\chi_1$ and $\chi_2$ are not functions, but just real
variables.
Here we use a mixed representation in terms of position variables
$Re\varphi$ and $Im\varphi$ and conjugate momenta

\begin{equation}
Re\pi=-i\frac{\partial}{\partial Re\varphi} \ ,\
Im\pi=-i\frac{\partial}{\partial Im\varphi}
\end{equation}

The Hamiltonian may be given by a rotationally symmetric 2 dimensional
oscillator because the two $\omega$'s are just opposite.
From eq.(\ref{4.7}) the coefficient of
$\frac{\partial^2}{\partial Im\varphi^2}$ is

\begin{equation}
\frac{-\omega}{2<s.p|s.p.>}
\end{equation}
and that of $(Im\varphi)^2$ is

\begin{equation}
\frac{1}{2}<s.p.|s.p.>\omega
\end{equation}
where $|s.p>$ denotes the single particle state.
These  coefficients are the same for both oscillators and
thus the Hamiltonian reads

\begin{eqnarray}
H&=&-\frac{1}{2}|\omega|\frac{\partial}{\partial\varphi}
	\frac{\partial}{\partial\varphi^\ast}+
	\frac{1}{2}|\omega|\varphi^\ast\varphi\nonumber\\
&=&\frac{1}{2}|\omega|\left(-
	\frac{\partial^2}{\partial Re\varphi^2}-
	\frac{\partial^2}{\partial Im\varphi^2}+
	Re\varphi^2+Im\varphi^2\right) \ \ \ .\nonumber
\end{eqnarray}
which is expressed in the mixed representation as

\begin{equation}
H=\frac{1}{2}|\omega|\left(-\frac{\partial^2}{\partial Re\varphi^2}+
	Re\varphi^2+Im\pi^2-
	\frac{\partial^2}{\partial Im\pi^2}\right) \ \ \ .
\end{equation}
We may express $H$ in terms of the real parameterization $\chi_1$ and
$\chi_2$ by using the relations

\begin{eqnarray}
Re\varphi&=&-(1+i)\chi_1-(1-i)\chi_2\nonumber \\
Im\pi&=&(1-i)\chi_1+(1+i)\chi_2 \ \ \ .\nonumber
\end{eqnarray}
It may be convenient to define

\begin{equation}
\chi_\pm=\sqrt{2}(\chi_2\pm\chi_1)
\end{equation}
so that the Hamiltonian can be simply expressed

\begin{equation}
H=\frac{1}{2}|\omega|\left(
	\frac{\partial^2}{\partial\chi^2_-}-\chi^2_- -
	\frac{\partial^2}{\partial\chi^2_+}+\chi^2_+\right) \ \ \ .
\end{equation}
The inner product takes the form

\begin{equation}
<\tilde{\psi_1}|\tilde{\psi_2}>=
	\int\tilde{\psi_1}(-\chi_-,\chi_+)^\ast
	\tilde{\psi_2}(\chi_-,\chi_+)d\chi_-d\chi_+
\end{equation}
where

\begin{equation}
\tilde{\psi_i}(\chi_-,\chi_-)=
\psi_i(-\sqrt{2}\chi_++i\sqrt{2}\chi_- , \sqrt{2}\chi_++i\sqrt{2}\chi_-)
\end{equation}

As to be expected the Hamiltonian turns out to be two uncoupled
harmonic oscillators expressed in terms of $\chi_-$and $\chi_+$.
The $\chi_+$ oscillator is usual one while $\chi_-$  has the
following two deviations :
one is that it has over all negative sign.
The other one is that in the definition of the inner product $-\chi_-$
instead of $\chi_-$ is used in the bra wave function.
This is equivalent to abandoning a parity operation
$\chi_-\rightarrow -\chi_-$ in the inner product.

The energy spectrum is made up from all combinations of a positive
contribution $|\omega|(n_++\frac{1}{2})$ with a negative one
$-|\omega|(n_-+\frac{1}{2})$ so that

\begin{equation}
E=|\omega|(n_+-n_-) \ \ \ .
\label{5.10}
\end{equation}
The norm square of these combination of the eigenstates are
$(-1)^{n_-{-1}}$
which equals to the parity under the $\chi_-$ parity operation
$\chi_-\rightarrow -\chi_-$  .

If we consider the single particle state, the charge or the number of
particles is given by

\begin{eqnarray}
Q&=&\frac{i}{4}\left\{
	\pi^+,\varphi\right\}-\frac{i}{4}\left\{\varphi^+, \pi\right\}
	\nonumber\\
&=&\frac{1}{2}\chi^2_+-\frac{1}{2}\frac{\partial^2}{\partial\chi^2_+}+
	\frac{1}{2}\chi^2_--\frac{1}{2}\frac{\partial^2}{\partial\chi^2_-}
	-1 \ \ \ .
\end{eqnarray}
This is simply a sum of two harmonic oscillator Hamiltonians with the
same unit frequency.
Thus the eigenvalue $Q'$ of $Q$ can only take positive integer or
zero.
For a given value $Q'$ that is number of particle in either of the
two states, the energy can vary from
$E=-|\omega|Q'$ to $E=|\omega|Q'$ in integer steps in $2|\omega|$ .
Thus we can put

\begin{equation}
n_-=Q', Q'-1, Q'-2, \cdots, 0
\end{equation}
in
the negative $\omega$ states while in the positive energy state we have

\begin{equation}
n_+=Q-n_- \  .
\end{equation}
So the energy eq.(\ref{5.10}) can be written as

\begin{equation}
E=|\omega|(Q-2n_-)
\label{5.14}
\end{equation}
which is illustrated in Fig.2(a).
By going to the convergence condition along the real axis we get the
usual theory with correct vacuum, see Fig.2(b).

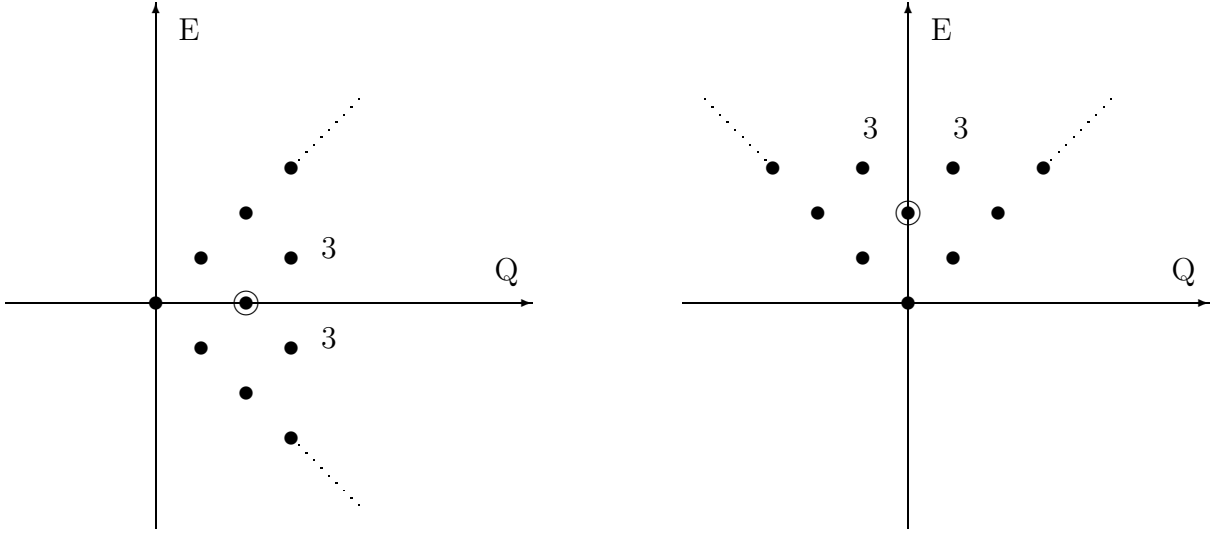
\begin{figure}[h]
\unitlength 1mm
\begin{picture}(150, 50)
\put(0,-30){\vector(0,1){70}}
\put(-20,0){\vector(1,0){70}}
\put(3,35){E}
\put(45,3){Q}
\put(0,0){\circle*{1.8}}
\put(6,6){\circle*{1.8}}
\put(12,12){\circle*{1.8}}
\put(18,18){\circle*{1.8}}
\put(6,-6){\circle*{1.8}}
\put(12,-12){\circle*{1.8}}
\put(18,-18){\circle*{1.8}}
\put(12,0){\circle{3}}
\put(12,0){\circle*{1.8}}
\put(18,6){\circle*{1.8}}
\put(18,-6){\circle*{1.8}}
\multiput(18,18)(1,1){10}{\line(0,0){0.1}}
\multiput(18,-18)(1,-1){10}{\line(0,0){0.1}}
\put(22,6){3}
\put(22,-6){3}
\put(100,-30){\vector(0,1){70}}
\put(70,0){\vector(1,0){70}}
\put(103,35){E}
\put(135,3){Q}
\put(100,0){\circle*{1.8}}
\put(106,6){\circle*{1.8}}
\put(112,12){\circle*{1.8}}
\put(118,18){\circle*{1.8}}
\put(94,6){\circle*{1.8}}
\put(88,12){\circle*{1.8}}
\put(82,18){\circle*{1.8}}
\put(100,12){\circle{3}}
\put(100,12){\circle*{1.8}}
\put(106,18){\circle*{1.8}}
\put(94,18){\circle*{1.8}}
\multiput(118,18)(1,1){10}{\line(0,0){0.1}}
\multiput(82,18)(-1,1){10}{\line(0,0){0.1}}
\put(106,22){3}
\put(94,22){3}
\end{picture}
\vskip40mm
(a) Naive vacuum
\hskip60mm
(b) True vacuum
\caption{charge vs energy in two state system }
{Energy $E$ versus charge $Q$=number of particles in the two state
system described in the text. 
This two state system is really a massive boson theory in 1 time+0
space dimensions.
The single dot $\cdot$ indicates that there is only one Fock space
state while the symbol $\bigodot$ means that there are two Fock space
states with the quantum number $E$ and $Q$.
The single dot with 3 is that there are 3 states and so on.
The triangles of dots are to be understood to extend to infinity.
The naive vacuum is depicted in Figure (a), while Figure (b) is for
the case of the true vacuum.}
\end{figure}

The wave function of the naive vacuum is given by

\begin{equation}
\psi_{n.\upsilon.}=
	N\exp\left(-\frac{1}{2}\chi^2_--\frac{1}{2}\chi^2_+\right)
\end{equation}
with a normalization constant $N$.
We may transform eq.(\ref{5.14}) in the mixed transformation back to the
position representation by Fourier transformation

\begin{eqnarray}
\psi_{n.\upsilon.}(Re\varphi, Im\varphi)&=&
	\int e^{iIm\Pi\cdot Im\varphi}\psi_{n.\upsilon}
	(Re\varphi, Im\Pi)dIm\Pi\nonumber\\
&=&N\int e^{iIm\Pi\cdot Im\varphi}e^{Im\Pi\cdot Re\varphi}dIm\Pi \ \ .
	\nonumber\\
&=&N\delta(Im\varphi-iRe\varphi) \ \ \ .
\end{eqnarray}
Here $\delta$-function is considered as a functional linear in test
functions that are analytic and goes down faster than any power in
real directions and no faster than a certain exponential in imaginary
direction.
This function may be called the distribution class $Z'$ according to
Gel'fand and Shilov \cite{gelfand}.
Thus our naive vacuum wave function is $\delta$-function that belongs
to $Z'$.

By acting polynomials in creation and annihilation operators to
the naive vacuum state we obtain the expression of the form

\begin{equation}
\sum_{n,m=0,1,\ldots}a_{n,m}(Re\varphi-iIm\varphi)^n
\delta^{(m)}(Re\varphi+Im\varphi) \ \ \ .
\label{5.17}
\end{equation}
Thus the wave functions of the double harmonic oscillator in the naive
vacuum world are composed of eq.(\ref{5.17}).

As long as the charge $Q$ is kept conserved, even an interaction term
such as an anharmonic double oscillator with phase rotation symmetry,
only the states of the form eq.(\ref{5.17})
can mix each other.
For such a finite quantum number $Q$ there is only a finite number of
these states of the form.
Therefore even to solve the anharmonic oscillator problem would be reduced to finite
matrix diagonalization.
In this sense the naive vacuum world is more easier to solve
than the correct vacuum world.

To higher dimensions we may extend our result of the double harmonic
oscillator for the naive vacuum.
The naive vacuum world would involve polynomials in the combinations
that are not present in the $\delta$-functionals and their
derivatives.

\section{The Naive vacuum world}

In this section we shall investigate properties of the naive vacuum
world.
It is obvious that this world has the following five inappropriate properties
from the point of view of phenomenological applications:

\begin{enumerate}
\renewcommand{\labelenumi}{\arabic{enumi})}
\item There is no bottom in the energy.

\item The Hilbert space is not a true Hilbert space because it is
not positive definite.
The states with an odd number of negative energy bosons get an extra
minus sign in the norm square.

We may introduce the boundary conditions to make a model complete which
may be different for negative energy states.
As will be shown
in the following paper ``Dirac Sear for Bosons II''[1]
in order to make an elegant CPT-like
symmetry we shall propose to take the boundary condition for the
negative energy states such that bound state wave functions blow up.

\item We cannot incorporate particles which are their own
antiparticles.
Thus we should think that all particles have some charges.

\item The naive vacuum world can be viewed as a quantum mechanical
system rather than second quantized field theory.
It is so because we think of a finite number of particles and the second
quantized naive vacuum world is in a superposition of various finite
numbers of particles.

\item As long as we accept the negative norm square there is no
reason for quantizing integer spin particles as bosons and half
integer ones as fermions.
Indeed we may find the various possibilities as is shown in table 1.
In this table we recognize that the wellknown spin-statistics theorem
is valid only under the requirement that the Hilbert space is positive
definite.
It should be noticed that in the naive vacuum world with integer spin
states negative norm squares exist anyway and so there is no
spin-statistics theorem.
When we go to the correct vacuum it becomes possible to avoid
negative norm square.
Then this calamity of indefinite Hilbert space is indeed avoided by
choosing the Bose or Fermi statistics according to the
spin-statistics theorem which is depicted in table 2.

\end{enumerate}

\begin{table}[h]
\begin{center}
\begin{tabular}{|c|c|c|}
\hline
\backslashbox{statistics}{spin}&
	$S=\frac{1}{2},\frac{3}{2},\ldots$&
	$S=0,1,\ldots$\\
\hline
Fermi-Dirac&$\|\cdots\|^2\geq 0$& Indefinite\\
\hline
Bose-Einstein&$\|\cdots\|^2\geq 0$ &Indefinite\\
\hline
\end{tabular}
\end{center}
\caption{Spin-statistics theorem for naive vacuum}
\end{table}

\vspace{2mm}
\begin{table}[h]
\begin{center}
\begin{tabular}{|c|c|c|}
\hline
\backslashbox{statistics}{spin}&
	$S=\frac{1}{2},\frac{3}{2},\ldots$&
	$S=0,1,\ldots$\\
\hline
Fermi-Dirac&$\|\cdots\|^2\geq 0$& Indefinite\\
\hline
Bose-Einstein&Indefinite&$\|\cdots\|^2\geq 0$ \\
\hline
\end{tabular}
\end{center}
\caption{Spin-statistics theorem for true vacuum}
\end{table}


\section{Conclusions}

We have put forward an attempt to extend also to bosons the idea of
Dirac sea for fermions.
We first consider one second quantization called the naive vacuum
world in which there exist a few positive and negative energy fermions
and bosons but no Dirac sea for fermions as well as bosons yet.
This first picture of the naive vacuum world model is very bad with
respect to physical properties in as far as no bottom in the
Hamiltonian.
For bosons this naive vacuum is even worse physically because in
addition to negative energies without bottom a state with an odd number
of negative energy bosons has negative norm square.
There is no real Hilbert space but only an indefinite one.
At this first step of the bosons the inner product for the Fock space
is not positive definite.
Thus this first step is completely out from the phenomenological point
of view for the bosons as well as for the fermions.
For the bosons there are for two major reasons : negative energy and
negative norm square.

However, from the point of view of theoretical study this naive vacuum
world at the first step is very attractive because the treatment of a
few particles is quantum mechanics rather than quantum field theory.
Furthermore by locality the system of several particles becomes free
in the neighborhood of almost all configurations except for the case
that some particles meet and interact.
We encourage the use of this theoretically attractive first stage as a
theoretical playground to gain physical understanding of the real world, the
second stage.

In the present article we studied the naive vacuum world at first
stage.
We would like to stress
the major results in
the following :

\begin{enumerate}
\renewcommand{\labelenumi}{\arabic{enumi})}
\item In the naive vacuum the single particles can be in position
eigenstates contrary to the particles in the ``true'' relativistic
theories.
\item The Fock space for the bosons is an indefinite one.
\item The bottom of the Hamiltonian is lost.
We made some detailed calculations on this issue.
\item We found the main feature of the wave functionals for the bosons.
They are derivatives of $\delta$-functionals of the complex field
multiplied by polynomials in the complex conjugate of the field.
These singular wave functionals form a closed class when acted upon by
polynomials in creation and annihilation operators.
Especially we worked through the case of one pair of a single particle
state with a certain momentum and the one with the opposite momentum.
\item In a subsequent ``Dirac Sea for Bosons II'' paper we will
present a CPT-like symmetry.
A reduced form of strong reflection provides an extra transformation
that is an analytic continuation of the wave function onto another
sheet among $2^{\frac{1}{2}N(N+1)}$ ones for the wave function of
the $N$ particle system.
The sheet structure occurs because $r_{ik}$ is a square root so that
it has 2 sheets.
For each of the $\frac{1}{2}N(N+1)$ pairs of particles there is a
dichotomic choice of sheet so that it gives $2^{\frac{1}{2}N(N+1)}$
sheets.
\end{enumerate}

The main point of our present work was to formulate the transition
from the naive vacuum of the first stage into the next stage of the
correct vacuum.
For fermions it is known to be done by filling the negative energy
states which is nothing but filling up the Dirac sea.
The corresponding procedure to bosons turns out to be that from each
negative energy single particle states one boson is removed, that is
minus one boson is added.
This removal cannot be done quite as physically as the adding of a
fermion, because there is a barrier to be crossed.

We studied this by using the harmonic oscillator corresponding to a
single particle boson state.
We replaced the usual Hilbert norm requirement of finiteness by the
requirement of analyticity of the wave function in the whole complex
$x$-plane except for $x=\pm\infty$.
The spectrum of this extended harmonic oscillator or the harmonic
oscillator with analytic wave function has an additional series of
levels with negative energy in addition to the usual one.
The wave functions with negative energy are of the form with Hermite
polynomials times $e^{\frac{1}{2}(\beta x)^2}$.

We note that there is a barrier between the usual states and the one
with negative excitation number because annihilation or creation
operators cannot cross the gap between these two sectors of states.
The removal of one particle from an empty negative energy state
implies crossing the barrier.
Although it cannot be done by a finite number of interactions
expressed as a polynomial in creation and annihilation operators we
may still think of doing that.
Precisely because of the barrier it is allowed to imagine the
possibilities that negative particle numbers could exist without
contradicting with experiment.

Once the barrier has been passed to the negative single particle
states in a formal way the model is locked in and those particles
cannot return to the positive states.
Therefore it is not serious that the correct vacuum for bosons get a
{\it higher} energy than the states with a positive or zero number of
particles in the negative energy ones.

Finally we mention that our deep motivation why we come to study in
detail already established standard 2nd quantization procedure in a
different point of view presented in this paper.
As was mentioned in the introduction once we come to the quantization
of the string theories we may face problem even in the 1st
quantization unless we use the light-cone gauge which was pointed out
long ago \cite{jackiw} by Jackiw et al. 
Furthermore there does not seem to exist satisfactory string field theories
except for Kaku-Kikkawa's light-cone string field theories.
We expect that our bosonic quantization procedure may clarify these
problem in the string theories.


\section*{Acknowledgment}
We would like to thank J. Greensite for useful discussions
on string field theories.
We also acknowledge R. Jackiw for telling us the ref [7].
Main part of this research was performed at YITP while one of us
(H.B.N.) stayed there as a visiting professor.
H.B.N. is grateful to YITP for hospitality during his stay.
M.N. acknowledges N.B.I for hospitality extended to him during his
visit.
This work is partially supported by Grants-in-Aid for Scientific
Research on Priority Areas, Number of Areas 763, ``Dynamics of Strings
and Fields'', from the Ministry of Education of Culture, Sports,
Science and Technology, Japan.


\end{document}